\documentclass[12pt]{iopart}
\usepackage{graphicx}

\usepackage{tikz}
\usetikzlibrary{cd}
\usetikzlibrary{calc}
\tikzset{anchorbase/.style={baseline={([yshift=-0.5ex]current bounding box.center)}},
  tinynodes/.style={font=\tiny, text height=0.25ex, text depth=0.05ex},
  smallnodes/.style={font=\scriptsize, text height=0.75ex, text depth=0.15ex},
  busual/.style={line width=1.0,color=blueberry},
  usual/.style={line width=2,color=black},
  rusual/.style={line width=1.0,color=dark-red,double},
  crossline/.style={preaction={draw=white,line width=6.0pt,-}},
  scrossline/.style={preaction={draw=white,line width=4.0pt,-}},
}

\definecolor{dark-red}{rgb}{0.7,0.25,0.25}

\usetikzlibrary{matrix, arrows, calc}
\newcommand{\ru}{to [out=0,in=270]}
\newcommand{\rd}{to [out=0,in=90]}
\newcommand{\ur}{to [out=90,in=180]}
\newcommand{\ul}{to [out=90,in=0]}
\newcommand{\lu}{to [out=180,in=270]}

\newcommand{\dr}{to [out=270,in=180]}
\newcommand{\dl}{to [out=270,in=0]}

\newcommand{\pr}{to [out=0,in=180]}

\newcommand{\genh}{3}

\newcommand{\su}[2]{
  \draw[usual,crossline] (#1,0) to (#1,\genh);
  \node at (#1,-.5) {\tiny #2};
}
\newcommand{\cupcap}[3]{
  \draw[usual] (#1,0) \ur (#1+.5,\genh/3) \rd (#1+1,0);
  \draw[usual] (#1,\genh) \dr (#1+.5,\genh-\genh/3) \ru (#1+1,\genh);
  \node at (#1,-.5) {\tiny #2};
  \node at (#1+1,-.5) {\tiny #3};
}

\newcommand{\longcap}[4]{
  \draw[usual] (#1,0) \ur (#1 + 1,\genh/3) \pr (#4,\genh/3) \rd (#4+0.5,\genh/4);  
  \draw[usual] (#1+1,0) \ur (#1+1.5,\genh/6) \pr (#4-0.5,\genh/6);
  \node at (#1,-.5) {\tiny #2};
  \node at (#1+1,-.5) {\tiny #3};
}

\newcommand{\longcup}[4]{
  \draw[usual] (#1,\genh) \dr (#1+1,\genh-\genh/3) \pr (#4,\genh-\genh/3) \ru (#4+0.5,\genh-\genh/4);
  \draw[usual] (#1+1,\genh) \dr (#1+1.5,\genh-\genh/6) \pr (#4-0.5,\genh-\genh/6);
}
\newcommand{\longcapp}[4]{
  \draw[usual,crossline]  (#4+0.5,\genh/4) \dl (#4,\genh/6) to (#4-0.5,\genh/6);
}

\newcommand{\longcupp}[4]{
  \draw[usual,crossline]  (#4+0.5,\genh-\genh/4) \ul (#4,\genh-\genh/6) to (#4-0.5,\genh-\genh/6);
}

 \newcommand{\longcupcap}[4]{
   \longcap{#1}{#2}{#3}{#4}
   \longcup{#1}{#2}{#3}{#4}
}

\newcommand{\longcupcapp}[4]{
  \longcapp{#1}{#2}{#3}{#4}
  \longcupp{#1}{#2}{#3}{#4}
}

\newcommand{\longwrap}[2]{
  \draw[usual] (#1+1,\genh/2) \dr (#1+1.5,\genh/6) \pr (#2-0.5,\genh/6);
  \draw[usual] (#1+1,\genh/2) \ur (#1+1.5,\genh-\genh/6) \pr (#2-0.5,\genh-\genh/6);
  \draw[usual,crossline] (#1,0) \ur (#1 + 1,\genh/3) \pr (#2,\genh/3) \rd (#2+0.5,\genh/4);  
  \draw[usual,crossline] (#1,\genh) \dr (#1+1,\genh-\genh/3) \pr (#2,\genh-\genh/3) \ru (#2+0.5,\genh-\genh/4);
}

\newcommand{\longwrapp}[2]{
  \longcupcapp{#1}{}{}{#2}
}

\newcommand{\longwrapb}[2]{
  \draw[usual] (#1,0) \ur (#1 + 1,\genh/3) \pr (#2,\genh/3) \rd (#2+0.5,\genh/4);  
  \draw[usual] (#1,\genh) \dr (#1+1,\genh-\genh/3) \pr (#2,\genh-\genh/3) \ru (#2+0.5,\genh-\genh/4);
  \draw[usual,crossline] (#1+1,\genh/2) \dr (#1+1.5,\genh/6) \pr (#2-0.5,\genh/6);
  \draw[usual,crossline] (#1+1,\genh/2) \ur (#1+1.5,\genh-\genh/6) \pr (#2-0.5,\genh-\genh/6);
}

\newcommand{\rsu}[2]{
\draw[rusual,crossline] (#1,0) to (#1,\genh);
\node at (#1,-.5) {\tiny #2};
}

\newcommand{\gens}[1]{
  \begin{tikzpicture}[anchorbase,scale=#1]
    \su{0}{1}
    \su{1}{2}
    \node at (2,\genh/2) {\tiny $\cdots$};
    \su{3}{~}
    \cupcap{4}{$j$}{$j$+1}
    \su{6}{~}
    \node at (7,\genh/2) {\tiny $\cdots$};
    \su{8}{$\ell$}
    \rsu{9}{}
  \end{tikzpicture}
}

\newcommand{\genss}[1]{
  \begin{tikzpicture}[anchorbase,scale=#1]
    \su{0}{1}
    \su{1}{2}
    \node at (2,\genh/2) {\tiny $\cdots$};
    \su{3}{~}
    \longcupcap{4}{$j$}{$j$+1}{9}    
    \su{6}{~}
    \node at (7,\genh/2) {\tiny $\cdots$};
    \su{8}{$\ell$}
   \node at (7,\genh/2) {\tiny $\cdots$};
    \rsu{9}{}
    \longcupcapp{4}{j}{j+1}{9}
  \end{tikzpicture}
}

\newcommand{\figtwo}{
  \[
  \begin{tikzpicture}[anchorbase,scale=.35]
    \su{0}{}
    \longcupcap{1}{}{}{3}    
    \rsu{3}{}
    \longcupcapp{1}{}{}{3}
    \begin{scope}[shift={(0,3)}]
      \longcupcap{0}{}{}{3}   
      \su{2}{} 
      \rsu{3}{}
      \longcupcapp{1}{}{}{3}
    \end{scope}    
    \begin{scope}[shift={(0,6)}]
      \su{0}{}
      \longcupcap{1}{}{}{3}    
      \rsu{3}{}
      \longcupcapp{1}{}{}{3}
      \end{scope}    
  \end{tikzpicture}
  \quad = \quad
  \begin{tikzpicture}[anchorbase,scale=.35]
    \su{0}{}
    \longcap{1}{}{}{3}    
    \rsu{3}{}
    \longcapp{1}{}{}{3}
    \begin{scope}[shift={(0,3)}]
      \longwrap{0}{3}   
      \rsu{3}{}
      \longwrapp{1}{3}
    \end{scope}    
    \begin{scope}[shift={(0,6)}]
      \su{0}{}
      \longcup{1}{}{}{3}    
      \rsu{3}{}
      \longcupp{1}{}{}{3}
      \end{scope}    
  \end{tikzpicture}
  \quad = \rho  \rho^{-1}
  \begin{tikzpicture}[anchorbase,scale=.35]
    \su{0}{}
    \longcap{1}{}{}{3}    
    \rsu{3}{}
    \longcapp{1}{}{}{3}
    \begin{scope}[shift={(0,3)}]
      \longwrapb{0}{3}   
      \rsu{3}{}
      \longwrapp{1}{3}
    \end{scope}    
    \begin{scope}[shift={(0,6)}]
      \su{0}{}
      \longcup{1}{}{}{3}    
      \rsu{3}{}
      \longcupp{1}{}{}{3}
      \end{scope}    
  \end{tikzpicture}
  \quad = \quad
  \begin{tikzpicture}[anchorbase,scale=.35]
    \su{0}{}
    \longcap{1}{}{}{3}    
    \rsu{3}{}
    \longcapp{1}{}{}{3}
    \begin{scope}[shift={(0,3)}]
      \su{0}{}
      \rsu{3}{}
    \end{scope}    
    \begin{scope}[shift={(0,6)}]
      \su{0}{}
      \longcup{1}{}{}{3}    
      \rsu{3}{}
      \longcupp{1}{}{}{3}
      \end{scope}    
  \end{tikzpicture}
  \]
}

\newcommand{\figthree}{
  \[
  \begin{tikzpicture}[anchorbase,scale=.35]
    \draw[usual,crossline] (0,\genh) to (0,2*\genh/3) \dr (1,\genh/3) \ru (1.5,\genh/2);
    \rsu{1}{}
    \draw[usual,crossline] (0,0) to (0,\genh/3) \ur (1,2*\genh/3) \rd (1.5,\genh/2);
  \end{tikzpicture}
  \quad = \quad \rho \;
  \begin{tikzpicture}[anchorbase,scale=.35]
    \draw[rusual,crossline] (1,\genh/2) to (1,\genh);
    \draw[usual,crossline] (0,0) to (0,\genh/3) \ur (1,2*\genh/3) \rd (1.5,\genh/2);
    \draw[usual,crossline] (0,\genh) to (0,2*\genh/3) \dr (1,\genh/3) \ru (1.5,\genh/2);
    \draw[rusual,crossline] (1,0) to (1,\genh/2);
  \end{tikzpicture}\qquad, \qquad
  \begin{tikzpicture}[anchorbase,scale=.35]
    \draw[rusual,crossline] (1,0) to (1,\genh/2);
    \draw[usual,crossline] (0,\genh) to (0,2*\genh/3) \dr (1,\genh/3) \ru (1.5,\genh/2);
    \draw[usual,crossline] (0,0) to (0,\genh/3) \ur (1,2*\genh/3) \rd (1.5,\genh/2);
    \draw[rusual,crossline] (1,\genh/2) to (1,\genh);
  \end{tikzpicture}
  \quad = \quad \rho \;
  \begin{tikzpicture}[anchorbase,scale=.35]
    \draw[usual,crossline] (0,0) to (0,\genh/3) \ur (1,2*\genh/3) \rd (1.5,\genh/2);
    \rsu{1}{}
    \draw[usual,crossline] (0,\genh) to (0,2*\genh/3) \dr (1,\genh/3) \ru (1.5,\genh/2);
  \end{tikzpicture}
  \]
}

\newcommand{\figfour}{
$
    \begin{tikzpicture}[anchorbase,scale=.35]
      \cupcap{0}{}{}
      \su{2}{}
      \su{3}{}
      \rsu{4}{}
      \node at (2,-1) {$e_1$};
    \end{tikzpicture}
    \qquad
    \begin{tikzpicture}[anchorbase,scale=.35]
      \su{0}{}
      \cupcap{1}{}{}
      \su{3}{}
      \rsu{4}{}
      \node at (2,-1) {$e_2$};
    \end{tikzpicture}
    \qquad
    \begin{tikzpicture}[anchorbase,scale=.35]
      \su{0}{}
      \su{1}{}
      \cupcap{2}{}{}
      \rsu{4}{}
      \node at (2,-1) {$e_3$};
    \end{tikzpicture}
    \qquad\begin{tikzpicture}[anchorbase,scale=.35]
      \longcupcap{0}{}{}{4}
      \su{2}{}
      \su{3}{}
      \rsu{4}{}
      \longcupcapp{0}{}{}{4}
      \node at (2,-1) {$f_1$};
    \end{tikzpicture}
    \qquad
    \begin{tikzpicture}[anchorbase,scale=.35]
      \su{0}{}
      \longcupcap{1}{}{}{4}
      \su{3}{}
      \rsu{4}{}
      \longcupcapp{1}{}{}{4}
      \node at (2,-1) {$f_2$};
    \end{tikzpicture}
    \qquad
    \begin{tikzpicture}[anchorbase,scale=.35]
      \su{0}{}
      \su{1}{}
      \longcupcap{2}{}{}{4}
      \rsu{4}{}
      \longcupcapp{2}{}{}{4}
      \node at (2,-1) {$f_3$};
    \end{tikzpicture}
$
\\
$
    \begin{tikzpicture}[anchorbase,scale=.35]
      \draw[usual,crossline] (0,\genh) \dr (0.5,2*\genh/3) \ru (1,\genh);
      \draw[usual,crossline] (2,\genh) \dr (2.5,2*\genh/3) \ru (3,\genh);
      \rsu{4}{}
      \node at (2,-.5) {$a_1$};
    \end{tikzpicture}
    \qquad
    \begin{tikzpicture}[anchorbase,scale=.35]
      \draw[usual,crossline] (0,\genh) \dr (1.5,\genh/3) \ru (3,\genh);
      \draw[usual,crossline] (1,\genh) \dr (1.5,2*\genh/3) \ru (2,\genh);
      \rsu{4}{}
      \node at (2,-.5) {$a_2$};
    \end{tikzpicture}
    \qquad
    \begin{tikzpicture}[anchorbase,scale=.35]
      \longcup{1}{}{}{4}
      \draw[usual,crossline] (0,\genh) to (0,2*\genh/3) \dr (2,\genh/6) \pr (4,\genh/6) \ru (4.5,\genh/4);
      \rsu{4}{}
      \draw[usual,crossline] (3,\genh) to (3,2*\genh/3) \dr  (4,\genh/3) \rd (4.5,\genh/4);
      \longcupp{1}{}{}{4}
      \node at (2,-.5) {$a_3$};
    \end{tikzpicture}
    \qquad    
    \begin{tikzpicture}[anchorbase,scale=.35]
      \longcup{0}{}{}{4}
      \draw[usual,crossline] (2,\genh) to (2,2*\genh/3) \dr  (3,\genh/3) to (3.8,\genh/3) \pr (4,\genh/3) \rd (4.5,\genh/4); 
      \draw[usual,crossline] (2.5,\genh/12) \ur (3.8,\genh/6) \pr  (4,\genh/6) \ru (4.5,\genh/4);
      \rsu{4}{}
      \draw[usual,crossline] (3,\genh) to (3,\genh/12) \dl (2.75,0) \lu (2.5,\genh/12);
      \draw[usual,crossline] (3.7,\genh/3) \pr (4,\genh/3) \rd (4.5,\genh/4);
      \longcupp{0}{}{}{4}
      \node at (2,-.5) {$a_4$};
    \end{tikzpicture}
    \qquad
    \begin{tikzpicture}[anchorbase,scale=.35]
      \longcup{2}{}{}{4}
      \draw[usual,crossline] (1,\genh) to (1,2*\genh/3) \dr  (3,\genh/3) to (4,\genh/3) \rd (4.5,\genh/4);
      \rsu{4}{}
      \draw[usual,crossline] (0,\genh) to (0,2*\genh/3) \dr  (4,\genh/6) \ru (4.5,\genh/4);
      \longcupp{2}{}{}{4}
      \node at (2,-.5) {$a_5$};
    \end{tikzpicture}
$
}

\begin{document}

\title[A coupled Temperley-Lieb algebra for the superintegrable chiral Potts chain]{A coupled Temperley-Lieb algebra 
for the superintegrable chiral Potts chain}

\author{Remy Adderton$^1$, Murray T. Batchelor$^{1,2,3}$ and Paul Wedrich$^{4}$
}

\address{$^1$ Department of Theoretical Physics, Research School of Physics, 
The Australian National University, Canberra ACT 2601, Australia}
\address{$^2$  Mathematical Sciences Institute, 
The Australian National University, Canberra ACT 2601, Australia}
\address{$^3$ Centre for Modern Physics, Chongqing University, Chongqing 40444, China}
\address{$^4$ Mathematical Sciences Research Institute, Berkeley CA 94720, USA}

\begin{abstract}
The hamiltonian of the $N$-state superintegrable chiral Potts (SICP) model is written in terms of a coupled algebra 
defined by $N-1$ types of Temperley-Lieb  generators. This generalises a previous result for $N=3$ 
obtained by J. F. Fjelstad and T. M\r{a}nsson [J. Phys. A {\bf 45} (2012) 155208]. 
A pictorial representation of a related coupled algebra is given for the $N=3$ case which involves a 
generalisation of the pictorial presentation 
of the Temperley-Lieb algebra to include a pole around which loops can become entangled. 
For the two known representations of this algebra, 
the $N=3$ SICP chain and the staggered spin-1/2 XX chain, 
closed (contractible) loops have weight $\sqrt{3}$ and weight $2$, respectively.
For both representations closed (non-contractible) loops around the pole have weight zero.
The pictorial representation provides a graphical interpretation of the algebraic relations.
A key ingredient in the resolution of diagrams is a crossing relation for loops encircling a pole which 
involves the parameter $\rho= e^{ 2\pi \mathrm{i}/3}$ for the SICP chain and $\rho=1$ for the staggered XX chain.
These $\rho$ values are derived assuming the Kauffman bracket skein relation. 
%
\end{abstract}
\maketitle

The $N$-state superintegrable chiral Potts (SICP) model, 
discovered by von Gehlen and Rittenberg \cite{vGR}, is defined on a chain of length $L$ by the Hamiltonian \cite{vGR,McCoy} 
\begin{equation}
H_{\mathrm{SICP}} = - \sum_{j=1}^L \sum_{n=1}^{N-1} \frac{2}{1-\omega^{-n}} \left[ \lambda \, \tau_j^n + (\sigma_j \sigma_{j+1}^\dagger)^n \right] .
\label{SICP}
\end{equation}
The parameter $\lambda$ is a temperature-like coupling and $\omega = e^{ 2\pi \mathrm{i}/N}$.
The  operators $\tau_j$ and $\sigma_j$ acting at site $j$ satisfy the relations 
\begin{equation}
\tau_j^\dagger = \tau_j^{N-1}, \qquad \sigma_j^\dagger = \sigma_j^{N-1}, \qquad \sigma_j \tau_j = \omega \, \tau_j \sigma_j, 
\end{equation}
with $\tau_j^N = \sigma_j^N =1$, where $1$ is the identity.
In terms of matrices, 
\begin{eqnarray}
\tau_j &=& 1 \otimes 1 \otimes  \cdots \otimes 1 \otimes \tau \otimes 1 \otimes \cdots \otimes 1,\\
\sigma_j &=& 1 \otimes 1 \otimes  \cdots \otimes 1 \otimes \sigma \otimes 1 \otimes \cdots \otimes 1,
\end{eqnarray}
where $1$ is the $N \times N$ identity matrix. The $N \times N$ shift and clock matrices $\tau$ and $\sigma$ are in position $j$, with 
\begin{equation}
\tau = 
\left( \begin{array}{ccccccc}
0 & 0 & 0 & \ldots & 0 & 1\\ 
1 & 0 & 0 & \ldots & 0 & 0\\ 
0 & 1 & 0 & \dots & 0 & 0\\
\vdots & \vdots & \vdots & & \vdots & \vdots\\
0 & 0 & 0 & \ldots & 1 & 0
\end{array} \right), 
\quad 
\sigma = 
\left( \begin{array}{ccccccc}
1 & 0 & 0 & \ldots & 0 & 0\\ 
0 & \omega & 0 & \ldots & 0 & 0\\ 
0 & 0 & \omega^2 & \dots & 0 & 0\\
\vdots &\vdots & \vdots & & \vdots & \vdots\\
0 & 0 & 0 & \ldots & 0 & \omega^{N-1}
\end{array} \right).
\end{equation}

The SICP chain is a special case of the more general chiral Potts model \cite{McCoy, Perk}. 
There has been a revival of interest in such models in the context of parafermionic edge zero modes and topological phases \cite{F,AF}.
The model defined by Hamiltonian (\ref{SICP}) is called superintegrable, because, 
beyond an infinite number of commuting conserved charges, it possesses additional symmetry 
generated by the Onsager algebra, owing to the Dolan-Grady condition \cite{DG} being satisfied. 
We remark that the SICP chain has been solved only for periodic boundary conditions, with the Onsager algebra playing a key role \cite{D}. 
In the above Hamiltonian, open boundary conditions are obtained by dropping the terms $(\sigma_L \, \sigma_{L+1}^\dagger )^n, n=1,2, \ldots, N-1$.
Here we will focus particularly on the case of open boundary conditions.

The other main ingredient for the present work is the Temperley-Lieb (TL) algebra \cite{TL}, also known as the Temperley-Lieb-Jones algebra \cite{J83}, 
which has enjoyed far reaching applications in both physics and mathematics. 
The TL algebra underpins a number of key models in statistical mechanics \cite{Baxter, Martin}, 
notably the spin-1/2 $XXZ$ and $N$-state Potts chains.
Both models can be written in terms of generators $e_j$ satisfying the TL algebra relations, from which their TL equivalence is established \cite{ABB}. 
Beyond the known representations in terms of spin operators, the TL algebra is arguably at its most powerful in the pictorial representation \cite{Martin, K, deGier}. 
Here we give the pictorial representation for an algebra related to the $N=3$
SICP chain, which involves two coupled copies of the TL algebra. 
The algebraic connection between the 3-state SICP chain and two coupled copies of the TL algebra has been established by Fjelstad and M\r{a}nsson \cite{FM}. 
Various other generalisations of the TL algebra are known, e.g., multi-coloured TL algebras \cite{GP,BJ,GM}.

For arbitrary $N$, we define generators $e_j^{(k)}$, $k=1,\ldots, N-1$, by
\begin{eqnarray}
e_{2j-1}^{(k)} &=& \frac{1}{\sqrt N} \sum_{n=1}^N (\omega^k \tau_j)^n, \label{def1} \\
~~ e_{2j}^{(k)} &=& \frac{1}{\sqrt N} \sum_{n=1}^N (\omega^k \sigma_j \sigma_{j+1}^\dagger)^n. \label{def2}
\end{eqnarray}
The relations satisfied by these generators include the coupled TL algebra relations
\begin{eqnarray}
~~~~~~ ( e_j^{(k)} )^2 &=& Q \, e_j^{(k)} \label{cTL1}\\
e_j^{(k)} e_{j\pm 1}^{(l)}  e_j^{(k)} & = & e_j^{(k)} \label{couple}\\
~~~~~~ e_i^{(k)} e_j^{(l)} &=& e_j^{(l)} e_i^{(k)}, \quad |i-j| > 1 \label{cTL3}\\
~~~~~~ e_j^{(k)} e_j^{(l)} &=& e_j^{(l)} e_j^{(k)} = 0, \quad k \ne l  \label{cTL4}
\end{eqnarray}
with $Q= \sqrt{N}$. 
This algebra is composed of $N-1$ copies of the TL algebra, coupled via the relations (\ref{couple}).
The above algebraic relations reduce to those for a single TL algebra when $N=2$, for which the relations (\ref{cTL4}) are not applicable.

Noting that the generators $e_j^{(k)}$ satisfy the inverse relations 
\begin{eqnarray}
~~~~~~ (\tau_j)^n &=& 1 - \frac{1}{\sqrt N} \sum_{k=1}^{N-1} (1 - \omega^{-k n})  e_{2j-1}^{(k)}, \\
( \sigma_j \sigma_{j+1}^\dagger)^n   &=& 1 - \frac{1}{\sqrt N} \sum_{k=1}^{N-1} (1 - \omega^{-k n})  e_{2j}^{(k)},
\end{eqnarray}
the SICP hamiltonian (\ref{SICP}) can be written in terms of the generators.
In this way, with periodic boundary conditions imposed, 
\begin{equation}
H_{\mathrm{SICP}} = -(N-1)L(\lambda+1) + \frac{2}{\sqrt N} \sum_{j=1}^L \sum_{k=1}^{N-1} (N-k) (\lambda e_{2j-1}^{(k)} + e_{2j}^{(k)}).
\end{equation}
In deriving this result we have made use of the relation (result (2.18) of \cite{Baxter09})
\begin{equation}
\sum_{n=1}^{N-1} \frac{2 \, \omega^{k n}}{1-\omega^{-n}} = N - 2k -1, \qquad 0 \le k \le N.
\end{equation}
Similarly for open boundary conditions
\begin{eqnarray}
H_{\mathrm{SICP}} &=& - (N-1) (L (\lambda +1) -1 )  + \frac{2}{\sqrt{N}} \sum_{j=1}^{L} \sum_{k=1}^{N-1} \lambda (N-k) e_{2j-1}^{(k)} \nonumber \\
&\phantom{=}& \qquad + \frac{2}{\sqrt{N}} \sum_{j=1}^{L-1} \sum_{k=1}^{N-1} (N-k) e_{2j}^{(k)}.
\label{open}
\end{eqnarray}

The generators $e_j^{(k)}$ satisfy additional relations beyond
(\ref{cTL1})-(\ref{cTL4}), which have been studied in the $N=3$ case in
\cite{FM}. Among them are cubic relations, which for $e_j=e_j^{(1)}$ and
$f_j=e_j^{(2)}$ take the form
\begin{eqnarray}
	f_{j} e_{j \pm 1} e_{j} &&= \pm \, \mathrm{i}  \left(\omega^{\mp 1} e_{j \pm 1} e_{j} - f_{j \pm 1} e_{j} \right) + \omega^{\pm 1} e_{j} \,,\label{cubic1} \\
	                                 &&= \pm \, \mathrm{i}  \left(\omega^{\mp 1} f_{j} e_{j \pm 1}  - f_{j} f_{j \pm 1}  \right) + \omega^{\pm 1} f_{j} \,,\label{cubic2}\\
	e_{j} e_{j \pm 1} f_{j} &&= \mp \, \mathrm{i}  \left(\omega^{\pm 1} e_{j \pm 1} f_{j} - f_{j \pm 1} f_{j} \right) + \omega^{\mp 1} f_{j} \,,\label{cubic3}\\
                                         &&= \mp \, \mathrm{i}  \left(\omega^{\pm 1} e_{j} e_{j \pm 1}  - e_{j} f_{j \pm 1}  \right) + \omega^{\mp 1} e_{j} \,,\label{cubic4}\\ 	
                                        f_{j} f_{j \pm 1} e_{j}&&= \omega^{\pm 1} f_{j} e_{j \pm 1} e_{j}\,,\quad 
                                         e_{j} f_{j \pm 1} f_{j}=\omega^{\mp 1}e_{j} e_{j \pm 1} f_{j}. \label{cubic5}
                                        \end{eqnarray}
In relations (\ref{cubic1})-(\ref{cubic4}) we have used the identity $(\omega - \omega^2)/\sqrt 3 = \mathrm{i}$.
Analogous cubic relations for general $N$ can be readily obtained by numerical observation, 
being similar combinations of one- and two-body terms. 
For example, for $N=4$ with $e_j=e_j^{(1)}$, $f_j=e_j^{(2)}$ and $g_j=e_j^{(3)}$, a typical cubic relation is of the type 
\begin{equation}
f_1 e_2 e_1 = \frac12 (1- \mathrm{i} ) e_2 e_1 -  \frac12 (1+ \mathrm{i} )  g_2 e_1 - \mathrm{i}  f_2 e_1 + \mathrm{i} \, e_1\,, 
\end{equation}
with $f_1 f_2 e_1 = \mathrm{i} f_1 e_2 e_1$. 
In general, relations of the type (\ref{cubic5}) involving the triple product $e_j^{(k)} e_{j\pm 1}^{(l)}  e_j^{(m)}$ obey
\begin{equation}
e_j^{(k)} e_{j\pm 1}^{(l)}  e_j^{(m)} = \omega^{\pm (m-k)} e_j^{(k)} e_{j\pm 1}^{(l+1)}  e_j^{(m)} \,. 
\end{equation}
These relations imply the result 
\begin{equation}
e_j^{(k)} e_{j\pm 1}^{(l)}  e_j^{(m)} = \omega^{\pm (m-k)(n-l)} e_j^{(k)} e_{j\pm 1}^{(n)}  e_j^{(m)} \,. 
\end{equation}
for all allowed values of $j,k,l,m,n$ and $N$.

Returning to the $N=3$ case, the abstract algebra with generators $e_i$ and
$f_j$ for $1\leq j\leq n$ and relations (\ref{cTL1})-(\ref{cTL4})
and (\ref{cubic1})-(\ref{cubic5}) was introduced in \cite{FM} as the special
case of a family of algebras $\tilde{\mathcal{A}}_{n}(\alpha)$ at
$\alpha=e^{-\pi i/6}$.

\begin{figure}[t]
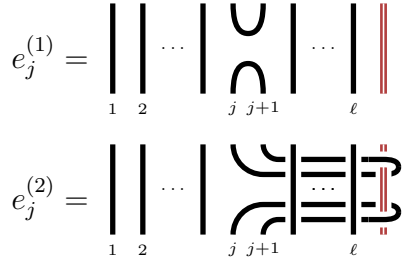

\begin{center}
\begin{eqnarray*}
    e_j^{(1)} &= \gens{.4} \\ e_j^{(2)} &= \genss{.4}
  \end{eqnarray*}
\caption{Pictorial representation of the generators $e_j^{(1)}$ and $e_j^{(2)}$.
} 
\label{fig1}
\end{center}
\end{figure}

\begin{figure}[htb]
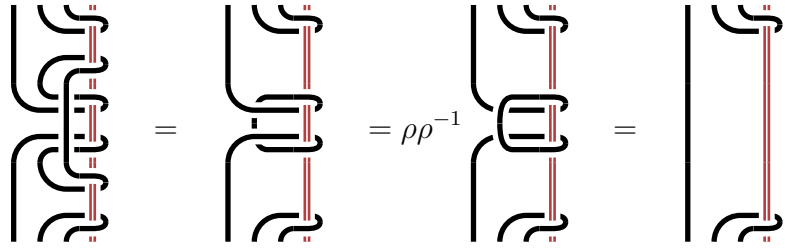

\begin{center}
\figtwo
\caption{Graphical version of the relation $f_2 f_1 f_2 = f_2$. 
The key ingredient in resolving the diagram is the crossing relations depicted in Figure \ref{fig3}.} 
\label{fig2}
\end{center}
\end{figure}

\begin{figure}[htb]
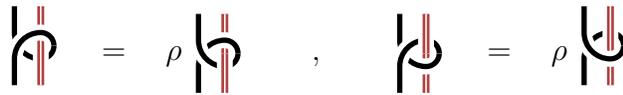

\begin{center}
\figthree
\caption{Crossing relations for a loop encircling a pole. For the SICP chain, for which $Q = \sqrt 3$, the parameter $\rho=\omega=e^{ 2\pi \mathrm{i}/3}$. 
%
}
\label{fig3}
\end{center}
\end{figure}

A pictorial representation of the generators $e_j$ and $f_j$ for this algebra is given in Figure \ref{fig1}.\footnote{A pictorial 
representation of an analogous algebra for general $N$, based on \cite{QW2}, will be discussed elsewhere.}
For a given $L$, we have odd diagrams $e_{2j-1}$ and $f_{2j-1}$ for $j=1,\ldots,L$ and even diagrams 
$e_{2j}$ and $f_{2j}$ for $j=1,\ldots,L-1$, 
and thus a total of $4L-2$ `cup-caps', running between $\ell=2L$ strands.
The key feature of the pictorial representation is a pole around which loops can become entangled.
Here we have chosen the position of the pole to be at one end of the chain.
In the associated loop diagrams, closed (contractible) loops have weight $Q$, with $Q=\sqrt 3$ in this example, 
while a closed (non-contractible) loop around any pole has weight zero.
With these definitions, each of the defining relations (\ref{cTL1})-(\ref{cTL4})
admits a graphical interpretation in terms of isotopies of diagrams and local
relations.
The graphical relations are similar to those for the single TL algebra \cite{K}. 
However, there is an interesting subtlety involved with the resolution of loops entangling a pole.
Consider the cubic relations defined by (\ref{couple}). 
%
%
%
The cubic relations of most interest are of the type $f_1 f_2 f_1 = f_1$ and $f_2 f_1 f_2 = f_2$.
The relation $f_2 f_1 f_2 = f_2$ is depicted in Figure \ref{fig2} along with the resolution of the diagram.
The key ingredient in the graphical interpretation of such relations is the identity shown in Figure \ref{fig3}.
The diagrammatic proof of $f_1 f_2 f_1 = f_1$ follows in similar fashion.
The proof of relations (\ref{cTL1}) and (\ref{cTL3}) is straightforward, while the 
proof of relations (\ref{cTL4}) relies on the vanishing of any closed loop encircling a pole. 

We point out that the case of one pole is also studied in the context of the `blob' algebra \cite{MS-blob}, a variation of the TL algebra 
which usually comes with two parameters: one corresponding to parameter $Q$ and one corresponding to the value of a closed loop  
encircling the pole or its equivalent (which we have set to zero). 
The case with $Q=-2$ and with loops encircling the pole also zero has been studied recently in the context of 
extremal weight projectors \cite{QW,QW2}.
Another variant is the one-boundary TL algebra \cite{dGN}.

The value of the parameter $\rho$ appearing in the crossing relations for loops
encircling a pole can be derived if one assumes the Kauffman bracket skein
relation between a crossing and its two planar resolutions \cite{K}. 
We set 
\begin{equation}
    \begin{tikzpicture}[fill opacity=.2,anchorbase,scale=.3]
      \draw[usual] (1,-1) to (-1,1);
      \draw[white, line width=.2cm] (1,1) to (-1,-1);
      \draw[usual] (1,1) to (-1,-1);
      \end{tikzpicture} ~:=~ 
    A  \;
        \begin{tikzpicture}[fill opacity=.2,anchorbase,scale=.3]
        \draw[usual] (1,1) to [out=225,in=315] (-1,1);
        \draw[usual] (1,-1) to [out=135,in=45] (-1,-1);
        \end{tikzpicture} 
        ~+~
        A^{-1}\;
        \begin{tikzpicture}[fill opacity=.2,anchorbase,scale=.3]
          \draw[usual] (1,1) to [out=225,in=135] (1,-1);
          \draw[usual] (-1,-1) to [out=45,in=315] (-1,1);
          \end{tikzpicture} ~,   
\end{equation}
and by a $\pi/2$ rotation, we also have
\begin{equation}
\begin{tikzpicture}[fill opacity=.2,anchorbase,scale=.3]
    \draw[usual] (1,1) to (-1,-1);
    \draw[white, line width=.2cm] (1,-1) to (-1,1);
    \draw[usual] (1,-1) to (-1,1);
    \end{tikzpicture} ~\phantom{:}=~ 
  A^{-1}  \;
      \begin{tikzpicture}[fill opacity=.2,anchorbase,scale=.3]
      \draw[usual] (1,1) to [out=225,in=315] (-1,1);
      \draw[usual] (1,-1) to [out=135,in=45] (-1,-1);
      \end{tikzpicture} 
      ~+~
      A\;
      \begin{tikzpicture}[fill opacity=.2,anchorbase,scale=.3]
        \draw[usual] (1,1) to [out=225,in=135] (1,-1);
        \draw[usual] (-1,-1) to [out=45,in=315] (-1,1);
        \end{tikzpicture} ~.  
\end{equation}
The Reidemeister 2 move holds iff $Q = - A^2 - A^{-2}$, where we recall that $Q$ is the weight or value of a closed, contractible, circle. 
This choice also ensures that the Reidemeister 3 move holds. 
Note that, generically there are four solutions $A\in \{\pm A_1,\pm A_2\}$.
In particular, for the values $Q= \sqrt{3}$ and $Q=2$ we have $A=e^{2\pi \mathrm{i} k/24}$ for $k\in \{5,7,17,19\}$ 
and $A\in\{ \mathrm{i},-\mathrm{i}\}$, respectively.
Finally, with regard to the Reidemeister 1 move, we have 
\begin{eqnarray}
  \begin{tikzpicture}[anchorbase,yscale=.35,xscale=-.35]
    \draw[usual,crossline] (0,0) to (0,\genh/3) \ur (1,2*\genh/3) \rd (1.5,\genh/2);
    \draw[usual,crossline] (0,\genh) to (0,2*\genh/3) \dr (1,\genh/3) \ru (1.5,\genh/2);
  \end{tikzpicture}
  ~ &=& ~
      -A^{-3}\;
      \begin{tikzpicture}[anchorbase, scale=.35]
        \draw [usual] (0,0) to (0,\genh);
        \end{tikzpicture} 
        ~ = ~
        \begin{tikzpicture}[anchorbase,scale=.35]
          \draw[usual,crossline] (0,\genh) to (0,2*\genh/3) \dr (1,\genh/3) \ru (1.5,\genh/2);
          \draw[usual,crossline] (0,0) to (0,\genh/3) \ur (1,2*\genh/3) \rd (1.5,\genh/2);
        \end{tikzpicture}        
        ~, \\
        \begin{tikzpicture}[anchorbase,xscale=-.35,yscale=.35]
          \draw[usual,crossline] (0,\genh) to (0,2*\genh/3) \dr (1,\genh/3) \ru (1.5,\genh/2);
          \draw[usual,crossline] (0,0) to (0,\genh/3) \ur (1,2*\genh/3) \rd (1.5,\genh/2);
        \end{tikzpicture}   
        ~ &=& ~
          -A^{3\phantom{-}}\;
          \begin{tikzpicture}[anchorbase, scale=.35]
            \draw [usual] (0,0) to (0,\genh);
            \end{tikzpicture}  
            ~ = ~
            \begin{tikzpicture}[anchorbase,scale=.35]
              \draw[usual,crossline] (0,0) to (0,\genh/3) \ur (1,2*\genh/3) \rd (1.5,\genh/2);
              \draw[usual,crossline] (0,\genh) to (0,2*\genh/3) \dr (1,\genh/3) \ru (1.5,\genh/2);
            \end{tikzpicture} ~.
          \end{eqnarray}

Now suppose that a circle wrapping around a red line is zero. It follows that 
\begin{eqnarray}  
  \begin{tikzpicture}[anchorbase,scale=.35]
    \draw[usual,crossline] (0,\genh) to (0,2*\genh/3) \dr (1,\genh/3) \ru (1.5,\genh/2);
    \rsu{1}{}
    \draw[usual,crossline] (0,0) to (0,\genh/3) \ur (1,2*\genh/3) \rd (1.5,\genh/2);
  \end{tikzpicture}
  ~ &=& ~
  A \;
  \begin{tikzpicture}[anchorbase,scale=.35]
    \draw[usual,crossline] (0,0) \ur (1,\genh/3) \ru (1.5,\genh/2);
    \rsu{1}{}
    \draw[usual,crossline] (0,\genh) \dr (1,2*\genh/3) \rd (1.5,\genh/2);
  \end{tikzpicture}
  + 0 ~, 
  \\
  \begin{tikzpicture}[anchorbase,scale=.35]
    \draw[rusual,crossline] (1,\genh/2) to (1,\genh);
    \draw[usual,crossline] (0,0) to (0,\genh/3) \ur (1,2*\genh/3) \rd (1.5,\genh/2);
    \draw[usual,crossline] (0,\genh) to (0,2*\genh/3) \dr (1,\genh/3) \ru (1.5,\genh/2);
    \draw[rusual,crossline] (1,0) to (1,\genh/2);
  \end{tikzpicture}
  ~ &=& ~
  0 + A^{-1} 
  \begin{tikzpicture}[anchorbase,scale=.35]
    \draw[usual,crossline] (0,0) \ur (1,\genh/3) \ru (1.5,\genh/2);
    \rsu{1}{}
    \draw[usual,crossline] (0,\genh) \dr (1,2*\genh/3) \rd (1.5,\genh/2);
  \end{tikzpicture} ~.
\end{eqnarray}
These diagrams agree up to a factor of $A^2$.
For oriented links, the symmetry between the crossing and the inverse of its $\pi/2$ rotation is broken. 
Thus one makes the following definition for the positive and negative oriented
crossings.
\begin{eqnarray}  
\begin{tikzpicture}[fill opacity=.2,anchorbase,scale=.3]
    \draw[usual,->] (1,-1) to (-1,1);
    \draw[white, line width=.2cm] (1,1) to (-1,-1);
    \draw[usual,<-] (1,1) to (-1,-1);
    \end{tikzpicture} ~&:=&~
  -A^{4\phantom{-}}  \;
      \begin{tikzpicture}[fill opacity=.2,anchorbase,scale=.3]
      \draw[usual] (1,1) to [out=225,in=315] (-1,1);
      \draw[usual] (1,-1) to [out=135,in=45] (-1,-1);
      \end{tikzpicture} 
      \; -\;
      A^{2\phantom{-}} \;
      \begin{tikzpicture}[fill opacity=.2,anchorbase,scale=.3]
        \draw[usual] (1,1) to [out=225,in=135] (1,-1);
        \draw[usual] (-1,-1) to [out=45,in=315] (-1,1);
        \end{tikzpicture} 
        ~, \\
        \begin{tikzpicture}[fill opacity=.2,anchorbase,scale=.3]
          \draw[usual,<-] (1,1) to (-1,-1);
          \draw[white, line width=.2cm] (1,-1) to (-1,1);
          \draw[usual,->] (1,-1) to (-1,1);
          \end{tikzpicture} 
          ~&:=&~ 
        -A^{-4}  \;
            \begin{tikzpicture}[fill opacity=.2,anchorbase,scale=.3]
            \draw[usual] (1,1) to [out=225,in=315] (-1,1);
            \draw[usual] (1,-1) to [out=135,in=45] (-1,-1);
            \end{tikzpicture} 
            \;-\;
            A^{-2}\;
            \begin{tikzpicture}[fill opacity=.2,anchorbase,scale=.3]
              \draw[usual] (1,1) to [out=225,in=135] (1,-1);
              \draw[usual] (-1,-1) to [out=45,in=315] (-1,1);
              \end{tikzpicture}   ~.
 \end{eqnarray}
Here we have simply rescaled the crossing by $-A^{\pm 3}$ depending on the sign of the oriented crossing. 
This makes them Reidemeister 1 invariant.
For the oriented crossing, we now have
\begin{eqnarray}  
    \begin{tikzpicture}[anchorbase,scale=.35]
      \draw[usual,crossline,<-] (0,\genh) to (0,2*\genh/3) \dr (1,\genh/3) \ru (1.5,\genh/2);
      \rsu{1}{}
      \draw[usual,crossline] (0,0) to (0,\genh/3) \ur (1,2*\genh/3) \rd (1.5,\genh/2);
    \end{tikzpicture}
    ~ &=& ~
    -A^4 \;
    \begin{tikzpicture}[anchorbase,scale=.35]
      \draw[usual,crossline] (0,0) \ur (1,\genh/3) \ru (1.5,\genh/2);
      \rsu{1}{}
      \draw[usual,crossline] (0,\genh) \dr (1,2*\genh/3) \rd (1.5,\genh/2);
    \end{tikzpicture}
    + 0 ~,
    \\
    \begin{tikzpicture}[anchorbase,scale=.35]
      \draw[rusual,crossline] (1,\genh/2) to (1,\genh);
      \draw[usual,crossline] (0,0) to (0,\genh/3) \ur (1,2*\genh/3) \rd (1.5,\genh/2);
      \draw[usual,crossline,<-] (0,\genh) to (0,2*\genh/3) \dr (1,\genh/3) \ru (1.5,\genh/2);
      \draw[rusual,crossline] (1,0) to (1,\genh/2);
    \end{tikzpicture}
    ~ &=& ~
    0 - A^{-4} 
    \begin{tikzpicture}[anchorbase,scale=.35]
      \draw[usual,crossline] (0,0) \ur (1,\genh/3) \ru (1.5,\genh/2);
      \rsu{1}{}
      \draw[usual,crossline] (0,\genh) \dr (1,2*\genh/3) \rd (1.5,\genh/2);
    \end{tikzpicture} ~.
 \end{eqnarray}
These diagrams agree up to a factor of $A^8$.
It follows that the prefactor $\rho$ defined in Figure 3 takes the values $\rho=\omega=e^{ 2\pi \mathrm{i}/3}$ for $Q = \sqrt 3$ 
and $\rho=1$ for $Q = 2$.

By using the above diagrammatic interpretation in terms of the Kauffman bracket
skein relation and the wrapping-around-a-pole relation, we have confirmed that
each of the additional cubic relations (\ref{cubic1})-(\ref{cubic5}) are also satisfied
diagrammatically. Thus, we obtain a pictorial representation of the algebra
$\tilde{\mathcal{A}}_{n}(\alpha)$ at $\alpha=e^{-\pi i/6}$. At this point we
note that the algebra elements (\ref{def1})-(\ref{def2}) satisfy additional
relations, some accounted for in the family of algebras
$\mathcal{A}_{n}(\alpha)$ from \cite{FM}, which do not hold in the pictorial
interpretation without imposing further relations.
%
%
%

Fjelstad and M\r{a}nsson \cite{FM} have also shown that the staggered spin-1/2 XX chain is associated with a representation of the
coupled algebra (\ref{cTL1})-(\ref{cTL4}) with two types of generators $e_j$ and $f_j$ and $Q=2$.
Here we have thus also given a pictorial representation for this model.
For the single pole case with $Q=2$, the value $\rho=1$ in the crossing relation for loops encircling a pole 
applies to the staggered spin-1/2 XX chain. 
This model was solved long ago via free fermions \cite{KT,PerkXX}. 
The generators are defined in terms of the Pauli spin matrices by \cite{FM}
\begin{eqnarray}
e_j &=& \frac{1}{2} (1 - \sigma^z_j \sigma^z_{j+1} + \sigma^x_j \sigma^x_{j+1} +\sigma^y_j \sigma^y_{j+1} ), \\
f_j &=& \frac{1}{2} (1 - \sigma^z_j \sigma^z_{j+1} - \sigma^x_j \sigma^x_{j+1} -\sigma^y_j \sigma^y_{j+1}).
\end{eqnarray}
The hamiltonian of the staggered XX chain, defined on a chain of $L$ sites with open boundary conditions, follows as
\begin{equation}
H_{\mathrm{sXX}} = \sum_{j=1}^{L-1} \lambda_j (e_j - f_j) , \label{sXX}
\end{equation}
where the values of the parameter $\lambda_j$ are staggered such that, 
e.g., $\lambda_j= \lambda$ for $j$ odd and $\lambda_j= 1$ for $j$ even.
This model is equivalent to the Su-Schrieffer-Heeger model of polyacetylene \cite{SSH}, 
recognised as a fundamental model of a topological insulator exhibiting edge states \cite{LXC,GKS}.

The cubic relations for the staggered XX chain have a similar structure to those for the SICP model.
We find  
\begin{eqnarray}
	f_{j} e_{j \pm 1} e_{j} &&=  e_{j \pm 1} e_{j} + f_{j \pm 1} e_{j}  -  e_{j} \,,\\
	                                 &&=  f_{j} e_{j \pm 1}  + f_{j} f_{j \pm 1} -  f_{j} \,,\\
	e_{j} e_{j \pm 1} f_{j} &&=  e_{j \pm 1} f_{j} + f_{j \pm 1} f_{j} - f_{j} \,,\\
                                         &&= e_{j} e_{j \pm 1}  + e_{j} f_{j \pm 1}  - e_{j} \,,  	
\end{eqnarray}
with the additional relations $f_{j} f_{j \pm 1} e_{j}= f_{j} e_{j \pm 1} e_{j}$ 
and $e_{j} f_{j \pm 1} f_{j}=e_{j} e_{j \pm 1} f_{j}$.
We have also checked that each of these relations is consistent with the diagrammatic interpretation.

\begin{figure}[htb]
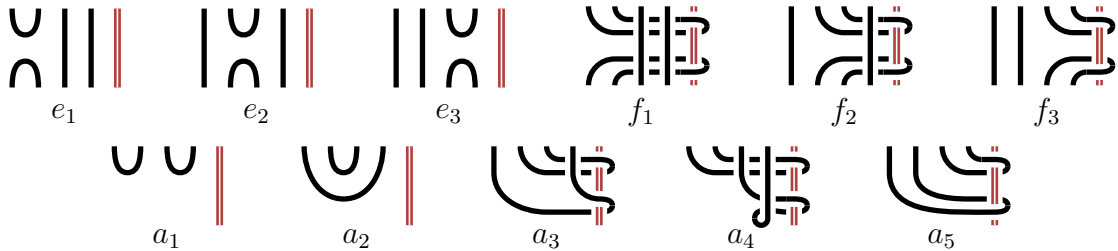

\begin{center}
\figfour
\caption{Generators and basis states for the ground state sector of the $N=3$ SICP chain for $L=2$. 
Note that the crossing in state $a_4$ can be resolved by using the first crossing relation depicted in Figure \ref{fig3}, 
thereby introducing a factor $\omega^{-1}$.}
\label{fig4}
\end{center}
\end{figure}

As an illustrative example of the algebraic/pictorial approach applied to the SICP chain for $N=3$, 
consider the simplest case $L=2$ with open boundary conditions.
Here $H_{\mathrm{SICP}}$ is a $9 \times 9$ matrix. 
The eigenvalues in the ground-state sector of the Hamiltonian are constructed from the words 
$a_1 = e_1 e_3 e_2$, $a_2 = e_2 a_1$, $a_3 = f_2 a_1$, $a_4 = f_1 a_2$ and $a_5 = f_3 a_2$. 
These five basis states, along with the six generators, are depicted in Figure \ref{fig4}.
In constructing the loop model hamiltonian a crucial point in resolving the related diagrams is to make use of the relations depicted in Figure \ref{fig3}.
In this way we have checked that the resulting $5 \times 5$ matrix recovers the corresponding five eigenvalues of $H_{\mathrm{SICP}}$, 
which for this sector, includes the groundstate eigenvalue. 
The remaining eigenvalues of $H_{\mathrm{SICP}}$ are obtained similarly.
We have also performed similar checks for larger values of $L$ and also for the staggered XX chain.
Such calculations are indeed very instructive, confirming in this way, e.g., the crossing relations depicted in Figure \ref{fig3}.

With the form of the staggered hamiltonian (\ref{sXX}) in view, we now reconsider the hamiltonian of the SICP model.  
It was shown for $N=3$ that $H_{\mathrm{SICP}}$ can be written \cite{FM}
\begin{equation}
H = -\frac{2}{\sqrt{3}} \sum_{j=1}^{L} \left[ \lambda \left( e_{2j-1} - f_{2j-1} \right) + \left(  e_{2j}- f_{2j}    \right) \right]. 
\label{Hef}
\end{equation}
Using the definitions (\ref{def1}) and (\ref{def2}), with again $(\omega - \omega^2)/{\sqrt3}= \mathrm{i}$, gives 
\begin{eqnarray}
e_{2j-1} - f_{2j-1} &=& \mathrm{i} \left(\tau_j - \tau_j^2 \right),  \\
~~~~~ e_{2j}- f_{2j} &=& \mathrm{i} \left[\sigma_j \sigma_{j+1}^\dagger - (\sigma_j \sigma_{j+1}^\dagger)^2 \right]. 
\end{eqnarray}
The hamiltonian (\ref{Hef}) thus becomes 
\begin{equation}
H = -\frac{2 \,\mathrm{i}}{\sqrt{3}} \sum_{j=1}^{L} \left[ \lambda \left(\tau_j - \tau_j^2 \right) + \sigma_j \sigma_{j+1}^\dagger - (\sigma_j \sigma_{j+1}^\dagger)^2 \right], 
\label{new}
\end{equation}
where the last terms $\sigma_L \sigma_{L+1}^\dagger$ and $(\sigma_L \sigma_{L+1}^\dagger)^2$ are omitted for open boundary conditions.
This hamiltonian has precisely the 
same eigenspectrum as $H_{\mathrm{SICP}}$ defined in (\ref{SICP}) with open boundary conditions.
%
%
We remark that (\ref{new}) appears to be a new way of writing the $N=3$ SICP hamiltonian, 
originating from the staggered nature of the coupled operators $e_j$ and $f_j$ between odd and even sites.

There are a number of immediate questions arising from this work, which we have begun to at least partially address. 
For example, it is known that the TL algebra, along with the pictorial representation, 
can be used to derive the full eigenspectrum of the TL Hamiltonian, 
in that case via the Bethe Ansatz \cite{L, MS, dGP, AK, NP}. 
The question then is if the SICP eigenspectrum can be obtained via the coupled TL algebra and pictorial representation given here, 
where we particularly have in mind a solution for open boundary conditions. 
Similarly one can also consider the periodic version.  
A related issue is finding other possible representations of the coupled algebra (\ref{cTL1})-(\ref{cTL4}). 
So far we know the SICP representation with $Q = \sqrt N$ for which there are $N-1$ coupled copies of the TL algebra. 
The other known representation is for the staggered spin-1/2 XX chain mentioned above, with $Q=2$ and two copies of the TL algebra.
It remains to be seen if this latter representation can be deformed or extended in some way for arbitrary $Q$, 
thereby establishing a TL-type equivalence with the $N=3$ SICP chain.
In concluding we also mention that the coupled TL algebra and its pictorial representation should  
in principle provide a pathway towards Baxterization \cite{Jones}, the process of adding a spectral parameter 
to algebraic relations to construct a solution of the Yang-Baxter equation, 
as originally achieved for TL, among other algebras. 
From the pictorial perspective, the inclusion of poles is expected to play a key role. 

\ack 

It is a pleasure to thank Rodney Baxter for his continual encouragement. 
This article is dedicated to his 80th birthday. 
This work has been supported by the Australian Research Council Discovery Project DP180101040 and by the   
National Natural Science Foundation of China Grant No. 11575037. P.W. was
supported by the National Science Foundation under Grant No. DMS-1440140, while
in residence at the Mathematical Sciences Research Institute in Berkeley,
California, during the Spring 2020 semester.

\end{document}